# A time-domain phase diagram of metastable states in a charge ordered quantum material

*Jan Ravnik[1,2], Michele Diego[1], Yaroslav Gerasimenko[1,2], Yevhenii Vaskivskyi[1], Igor Vaskivskyi[3], Tomaz Mertelj[1,3], Jaka Vodeb[1] and Dragan Mihailovic[\*1,3,4]*

[1]Complex Matter Department, Jozef Stefan Institute, Jamova 39, SI-1000 Ljubljana, Slovenia, E-mail: dragan.mihailovic@ijs.si
[2]Laboratory for Micro and Nanotechnology, Paul Scherrer Institut, Forschungsstrasse 111, 5232 Villigen PSI, Switzerland
[3]Center of Excellence on Nanoscience and Nanotechnology – Nanocenter (CENN Nanocenter), Jamova 39, SI-1000 Ljubljana, Slovenia
[4]Department of Physics, Faculty of Mathematics and Physics, University of Ljubljana, Jadranska 19, SI-1000 Ljubljana, Slovenia

**Abstract**

Metastable self-organized electronic states in quantum materials are of fundamental importance, displaying emergent dynamical properties that may be used in new generations of sensors and memory devices. Such states are typically formed through phase transitions under non-equilibrium conditions and the final state is reached through processes that span a large range of timescales. By using time-resolved optical techniques and femtosecond-pulse-excited scanning tunneling microscopy (STM), the evolution of the metastable states in the quasi-two-dimensional dichalcogenide 1$T$-TaS$_2$ is mapped out on a temporal phase diagram using the photon density and temperature as control parameters on timescales ranging from $10^{-12}$ to $10^3$ s. The introduction of a time-domain axis in the phase diagram enables us to follow the evolution of metastable emergent states created by different phase transition mechanisms on different timescales, thus enabling comparison with theoretical predictions of the phase diagram and opening the way to understanding of the complex ordering processes in metastable materials.

Short optical or electrical pulses can create non-equilibrium conditions that lead to electronic self-organization in quantum materials which can be usefully applied in various devices[1–9]. Recent rapid progress in the time-domain investigations of non-equilibrium phase transitions has led to the observation of a variety of emergent transient and metastable states in complex quantum materials, including organic electronic crystals[10], oxides[11,12], dichalcogenides[13], and fullerene superconductors[14]. Self-organized non-equilibrium matter that emerges after pulsed laser excitation evolves in a sequence of processes that eventually cause it to reach the ground state. However, sometimes new emergent metastable states may be created during the relaxation process which are not present in the equilibrium phase diagram[13].

The formation of such metastable states can occur by different mechanisms: in addition to first or second order transitions characterized along the Ehrenfest scheme, topological and jamming transitions may occur under non-equilibrium conditions –leading to fundamentally different dynamical ordering phenomena and diverse spatial orders. For example, domain structures can emerge through the Kibble-Zurek[15,16] mechanism in second-order transitions or transient mesoscopic phase separation in first-order transitions. To obtain broader understanding of the dynamics of mesoscopic emergent phenomena that accompany such phase transitions the boundaries between metastable phases need to be determined on different timescales.

Presenting phase diagrams of *temporally-evolving* systems is a significant challenge. Most experimental techniques that can give sufficiently detailed microscopic and mesoscopic structure, such as electron and scanning probe microscopies, work well only on timescales $10^{-3} \sim 10^3$ s. On the other hand, the investigation of ultrashort timescale phenomena that occur in phase transition require stroboscopic repetitive scanning, which requires that the system dynamics is periodic, i.e. the transition outcome should be the same each time, so that it can be probed many times at the same moment in its cycle, and the signal can be averaged over many cycles. This has significant limitations, such as the requirement that all phenomena need to fully relax between measurements. The outcomes of singular ultrafast events, such as ultrafast phase transitions caused by a single laser shot that lead to metastable states which are of particular recent interest here, are an even greater challenge. Single shot techniques are limited by balancing signal to-noise and damage by the probe unless the metastable state lifetime is sufficiently long, so that each outcome can be examined by STM or AFM for example.

Yet, as we show here, using a combination of multi-pulse femtosecond time-resolved coherent phonon spectroscopy in combination with STM, a temporal phase diagram may be pieced together when the lifetime of the states can be tuned by temperature. The different phases that appear in the experiment on different timescales can then be compared with the results of a theoretical charged-lattice-gas model calculation of equilibrium and metastable photoinduced phases.

The studied material is a prototypical quasi-2D dichalcogenide $1T$-TaS$_2$, which was shown recently to display multiple metastable electronic and structural ordering phenomena. Its' phase diagram includes different structural polytypes ($1T$-TaS$_2$, $2H$-TaS$_2$ etc)[17], different configurational charge-ordered states[13,18–21], superconductivity[22] and a quantum spin liquid candidate state[23]. The $1T$ polytype is metallic above 550 K. In the range 350∼550 K it displays an incommensurate (IC) charge density wave (CDW), which undergoes a transition to a nearly-commensurate (NC) phase below ∼ 350 K. Below 180 K, the material becomes insulating and fully commensurate (C), discussed either in terms of a CCDW, a polaronic Wigner crystal[24–26] or a Mott state[22]. Upon heating, the material goes through a triclinic domain state in the range 220∼280 K, whereupon it reverts to the NC state. Among the many different equilibrium orders, the material also exhibits long-lived metastable states. Of particular interest are (i) the non-equilibrium topological transition to a 'hidden' (H) metallic state with chiral domain structures[19], and (ii) the jamming transition to an amorphous (A) state with hyperuniform electronic order[18], both of which can be reached by photoexcitation.

The non-equilibrium phases in 1$T$-TaS$_2$ were previously investigated by optical pump-probe techniques[27], transport measurements [1,3,4,28], time-resolved X-ray diffraction[29–31], and time-resolved electron diffraction (TrED)[32–35], sketching the timeline of the transient phenomena. After the creation of *e-h* pairs by the incident laser photons (in < 1 fs), melting of the Mott state takes place within ∼50 fs, while the periodic lattice modulation melts on the timescale of the collective mode 1/2-period (~200 fs)[36]. The transition to the H state was observed by coherent phonon measurements to take place in 450 fs[27]. TrED measurements[32] revealed a number of transient phases on the picosecond timescale dependent on photon density, with a semi-continuous rotation of the CDW ordering wavevector with respect to the crystal axes as a function of laser fluence. The ordering wavevector angles observed in TrED are consistent with the angles obtained from the Fourier transformed STM images[19] and with recent static X-ray measurements of the H state[37]. X-ray diffraction showed electronic domain fusion dynamics associated with the formation of the IC phase, described in terms of coherent domain formation on the picosecond timescale[35] followed by diffusive processes on a time-scale of ∼100 ps[31].

Importantly for this study, the lifetime of the hidden (H) state is temperature-dependent, ranging from an extrapolated ≫ $10^{10}$ s at 4 K to < $10^{-4}$ s at 150 K, which allows us to tune its relaxation time[2], facilitating stroboscopic measurements. The A phase was suggested to be stable to temperatures above 200 K[18] but the temperature-dependence of it's lifetime is not known.

**Results**

To map out the temporal phase diagram we link STM images of different phases with the femtosecond optical spectroscopy results. First, we identify which long-lived nonequilibrium phases we can reach with different photoexcitation densities. To this end, we conduct STM experiments at low temperatures (4 K), where all of the previously known photoinduced phases can be considered stable, and we can distinguish them by their distinct spatial electronic ordering. Next, we link the latter to specific coherent phonon spectral features measured with the low-fluence two-pulse transient reflectivity spectroscopy. With the phonon fingerprints of the various states ascertained, we present three pulse technique measurements designed to track the non-equilibrium phonon evolution on short timescales.

**STM experiments.** We use an ultrahigh vacuum (UHV) low temperature STM (LT Nanoprobe, ScientaOmicron) with optical access (Figure 1a). The 1$T$-TaS$_2$ samples are excited with an external laser. Since the STM scans cover a very small (30 × 30 nm$^2$) area of the elliptical Gaussian laser spot (~ 100 × 150 μm$^2$ at FWHM) we can investigate the effect of different laser fluences simply by choosing an appropriate area of the STM scan with respect to the center of the Gaussian beam. The excitation is done with a laser pulse with the peak fluence of 12 mJ/cm$^2$, which gives us a possibility to perform STM measurements over a wide range of fluences $F$ = 0~12 mJ/cm$^2$ (Figure 1a). Typical experiments are done with a single laser pulse photoexcitation, but multiple pulse excitations were also performed to understand the effects of heating the sample. The outcome of different 1 and 10$^6$ shot experiments are shown in Figures 1c and 1d, where the colors of triangles represent the different observed states (Figures 1e-j). In both cases, the system returns to the original (C) state after photoexcitation (Figure 1e, blue in 1c and 1d) in the region where the fluence $F$ < 1.5 mJ/cm$^2$. Above ~1.5 mJ/cm$^2$ the characteristic H state domain mosaics[19] consistently appear (Figure 1f, yellow in 1c and 1d), independent on the number of the excitation pulses. Increasing the fluence beyond ~ 3.5 mJ/cm$^2$ patches of the amorphous A state with a characteristic hyperuniform electronic structure[18] start to appear within the H-state areas (Figure 1g, pink in 1c and 1d). These appear very inconsistently, and independent of the number of excitation pulses. At even higher fluences, we observe irreversible structural changes (ISC) that cannot be annealed by heating the sample (black). These include single layer polytype transformations from 1T to 1H (Figure 1h), which appear as triangular patches of various sizes without charge ordering above 70 K[17]. Another form of ISC is the layer peel-off and the creation of 1D nanotube-like objects (Figure 1i). Eventually melting/ablation of the material is observed and an uneven surface is created with nano-scale modulations (Figure 1j) or craters a few microns in diameter (Figure 1b). Large regions of ISC sometimes appear in single pulse experiments, but are more common with pulse-train excitation (Figure 1d) (Figure 1c shows an example of a single pulse excitation, where ISC did not appear, however this is not always the case), implying that the accumulated heating of multiple pulses enhances the formation of ISC. The ISC areas are surrounded and sometimes intermixed either by H or A state on micrometer scale regions. We further note that for high fluences > 3 mJ/cm$^2$ the transition outcome is not perfectly reproducible on different areas of the sample. We attribute this to uneven thermal coupling between the sample and the substrate, possible imperfections of the layer stacking, or strains caused during cleaving.

**Femtosecond laser spectroscopy.** Having established the boundaries of the phase diagram on the STM timescale of ~10$^3$ s, we use femtosecond spectroscopy with three different pulse sequences (Figure 2a) to investigate the trajectory of the system on the timescale down to 10$^{-13}$ s via coherent phonons: (i) A single "Driving" (D) pulse causes the change of state, followed by a standard stroboscopic two-pulse Pump (P) - probe (p) experiment, in which the timescale is defined by the total time of the measurement (~10 minutes). The weakly-perturbative P-p protocol allows us to extract the near-equilibrium phonon fingerprints of the metastable phases on the same timescales and temperatures as in the STM experiments, thus matching the signatures obtained by the two techniques. (ii) A repetitive three-pulse D-P-p sequence gives access to the picosecond timescale. Here, each scan of the P-p experiment is taken with a small (ps) fixed delay between the D and P pulses. This is applicable for mapping the H state at temperatures where its lifetime is shorter than the repetition time (1 ms using a 1 kHz laser system) and thus the sample fully relaxes before the next D-P-p sequence arrives. (iii) The D-P-p technique, with the changed order of the pulses. Here the P-p pulses come a few tens of ps before the D pulse, which effectively produces ~ 1ms D-P delay (defined by the repetition rate). This way, we are able to establish whether the sample has relaxed between the pulses or in the case that it has not relaxed, we are able to see which state was present after 1 ms. To ascertain that no permanent change has occurred in the sample in the case when it does not completely relax in 1 ms, we turn off the D pulse and re-measure the reflectivity transient using only the weakly perturbative P-p sequence.

The C and H states can be distinguished by the frequency of the collective amplitude mode (AM)[13]. While the AM frequencies in both states are temperature dependent, the AM frequency of the H state is about 0.1 THz lower than the AM frequency of the C state at any given temperature[27], thus making the two states easy to distinguish. The T and NC states show a characteristic double peak in the phonon spectrum with lower amplitude and about 0.1 THz lower frequency than the H state at any temperature and can thus also be clearly differentiated[27]. In Figure 2a and b we show the transient reflectivity oscillations and the respective Fourier spectra for C, H and T states. A detailed analysis of the C, H, NC and T state AM peak frequencies and line shapes with multi phonon fits and their temperature dependences are given in Reference 27.

The data for different D pulse fluences $F = 0.2 \sim 10$ mJ/cm$^2$ were taken at 80, 100, 140, 160 and 200 K. In Figure 2c and d we show the transient reflectivity oscillations and the respective Fourier spectrum at 160 K at different D fluences after 30 picoseconds and after a millisecond after switching to the metastable state.

Following the difference between the AM peaks in the C and H state, the C/H boundary on the low-fluence side can be defined with a high degree of certainty. The photoinduced H state at fluences slightly above 1 mJ/cm$^2$ forms already on the picosecond timescale and relaxes at this temperature within the ~1 ms time between the pulses. The same was observed at 100 K and 140 K. At 80 K, the H state does not completely relax between the succesive pulses, which is in agreement with the STM data at 77 K.

On the high-fluence side, the boundary of the H state cannot be as clearly ascertained. With increasing fluence, we see lowering of the AM peak intensity, broadening of the peak and a further shift to lower frequencies. In this regime, we cannot easily recognize a unique fingerprint of any of the known states. To ascertain the presence of the A state, we compare the spectra of pristine and exposed samples 1 ms after photoexcitation at high temperatures, where the contribution of H phase is absent due to its short lifetime. We see that for fluences above ~3 mJ/cm$^2$ the sample does not relax completely to the C state in 1 ms, which is expected for the A state, but may also appear when other unidentified long-lived disordered phases or phase separation are present. As observed by STM for $F < 7$ mJ/cm$^2$, such transient states with no characteristic oscillation fingerprint eventually evolve into either the H or the A state. Since (i) multiple pulse experiments in STM show little or no difference from single pulse experiments (with fluences below the damage threshold) and (ii) repetitive single-shot measurements in the same spot in STM show different electronic ordering pattern of the H and A states after each pulse, we conclude that the final photoinduced states are 'reconfigured' by each pulse and independent of the initial electronic order.

For $F > 7$ mJ/cm$^2$, the AM peak intensity decreases even further, indicating the appearance of ISC. The irreversibility of the excitation process was tested by turning off the D pulse and measuring a control P-p transient from the same spot. At fluences up to 7 mJ/cm$^2$, the signal mostly recovers. With increasing the fluence beyond 7 mJ/cm$^2$, the signal recovers only partially, while finally, at fluences above 10 mJ/cm$^2$, the sample suffers enough damage to completely suppress the signal, which is consistent with the STM scans.

**The time-domain phase diagram** with a compilation of the transition outcomes for different $F$ and $T$ is shown in Figure 3, combining STM (triangles) and transient reflectivity data (squares). The time-axis signifies the time after the photoexcitation pulse (Figure 3a) grouping the data into three timescales: ultrashort ($< 10^{-10}$ s), intermediate ($10^{-3}$ s) and long ($10^3$ s). The $F$ axis is converted into the photon density taking a penetration depth of 30 nm at 800 nm (Figure 3b). The color shade (blue, yellow, pink, black) represent different states as shown in Figure 1e) – j). White represents unidentified transient states, or an inhomogeneous mixture of states, which cannot be separated spectroscopically.

We see that the H state has a nearly temperature-independent threshold fluence and is stable at low temperature in agreement with Ref. 3. When the measurement timescale is longer than the relaxation time, the H state disappears, thus the C/H boundary moves to lower temperature with time. The phase boundary of the C state observed by TRED[32] agrees remarkably well with the present measurements on short timescale. Also, the suppression of the CDW diffraction peaks is in agreement with the decrease of the AM peak intensity in the transient optical spectroscopy data reported here[27,32]. Finally, we note that the H state boundaries on long timescales are also consistent with the recent XRD measurements[37]. Comparing the appearance of the H state on short and long timescales we see that the H state immediately appears at low fluences, but it takes much longer to stabilize at higher fluences. On the other hand, on the long timescale at 77 K the H state is only visible at fluences above ~5 mJ/cm$^2$, but is absent at lower fluences, even though it was observed there on a short timescale. This implies that the relaxation first happens on the outskirts of the beam, where the excitation fluence is lower.

To obtain insight into the origin of different phases we compare the observed experimental phases on the STM timescale with the equilibrium configurational states obtained from theoretical treatment. The model considers the ordering of electrons subject to screened Coulomb interaction on a triangular atomic lattice, and was previously successfully applied to describe both irregular domain patterns[24,25] and hyperuniform polaron orders[18,24] in the H and A states, respectively. It's predictions can be compared with the experiment by assuming a correspondence between the photoexcited carrier density (which is proportional to incident photon density) and electron filling. The model defines the filling of the system as the number of electrons at the Fermi level divided by the number of atoms[24]. In 1T-TaS$_2$ with one electron per Ta atom, a $\sqrt{13} \times \sqrt{13}$ reconstruction of the CCDW state gaps 12 out of 13 electrons, resulting in 1/13 filling by the remaining electron[38,39]. Doping is defined as a change of the filling with respect to 1/13. Experimentally, the filling is obtained by counting the number of polarons per unit area in an STM image (1 polaron equals 1 electron) [18,24].

Monte-Carlo simulations using this model give a theoretical phase diagram (Figure 3c) that is consistent with the experimentally observed C (1/13 filling) and H states at ~ 4 % nominal doping [24,25], observed at the experimental photodoping of 0.09 photons/unit cell. Remarkably, they also predict the A state towards 1/11 filling (at ≳ 15 % nominal doping, observed at a threshold of ~0.3 photons/unit cell) [18,24]. Simulations also predict the existence of a uniform ordered state with 1/12 filling and close to 1/12 domain states around it, but these states are not observed experimentally. Possibly, the 1/12 superlattice is less stable and may only exist as a transient phase, but so far it has not been observed. In spite of the crudeness of the simple model, which correctly predicts the observed metastable charge configurations, we cannot expect it to predict their stability, for which one needs to consider additional effects, such as long-range order and topological defects created in the transition[19]. Larger density of such defects created by higher fluences may be the cause of the higher long-term stability of the metastable H phase at higher fluences, which was observed at 77 K. The amorphous state on the other hand is stabilized by the constraints imposed by jamming, which is an entirely different stabilization mechanism[18].

We conclude that by a carefully chosen combination of techniques, phase-diagram 'snapshots' can be obtained during the relaxation trajectory of a non-equilibrium system. Comparison of the experimental phase diagram with the theory under the assumption that photoexcitation is equivalent to doping confirms that three out of four phases can be reached with photoexcitation fluence as the only control parameter. While the transition outcome reproducibility is excellent on the low-fluence side (up to ~3 mJ/cm$^2$), various factors that are not under direct control contribute to variable transition outcomes on the high fluence side. This has important implications for the stability and potential device reliability. The understanding of the time-evolution of different nonequilibrium states revealed by temporal phase diagrams opens the way to the development of new functionalities in metastable quantum materials based on configurational electron ordering.

**Methods**

**Sample preparation.** 1$T$–TaS$_2$ crystals were grown by chemical transport method with iodine as a transport agent. The samples have average dimensions of 2 × 2 × 0.1 mm. For optical measurements, the samples were glued to a copper plate with thermally conductive vacuum glue and cleaved just before inserting into the cryostat. For STM measurements, the samples were glued to the holder with a UHV compatible silver paste and inserted into the machine. The samples were cleaved inside the UHV chamber to prevent surface contamination.

**STM experiments.** We use a low temperature STM (LT Nanoprobe, ScientaOmicron) with optical access. For the photoexcitation of the samples, we use a 100 kHz, 800 nm laser system, with 50 fs pulses. The number of shots was selected using an acousto-optic modulator. The laser pulses are guided into the STM chamber using an automatic stabilizing system with the positioning precision of < 5 μm. The laser beam profile was carefully determined externally with a CCD camera. A scanning electron microscope mounted above the STM allowed for precise tip positioning. The center position of the laser beam was determined from the perimeter defined by the border between the equilibrium and switched states.

**Femtosecond spectroscopy.** For the optical experiments, we use a 1 kHz, 800 nm laser system with 50 fs pulses. The sample is held in a vacuum cryostat with optical access. As an addition to the standard pump (P) and probe (p) pulses, we use the third driving (D) pulse to drive the sample out of equilibrium. We varied the fluence of the D pulse from 0.2 mJ cm$^{-2}$ to 10 mJ cm$^{-2}$. The fluences of the P and Pr pulses were always below 0.1 mJ cm$^{-2}$ to ensure minimal disturbance. We use a mechanical chopper to modulate the P pulse, while the D and p pulses are unmodulated. This makes the D pulse »invisible« in the repetitive measurement with the lock-in technique and we only observe its effect on the sample.

**Theoretical model.** The model is based on the charged lattice gas (CLG) Hamiltonian $H = \sum_{i,j} V(i,j) n_i n_j$ where $n_i$ is the occupational number of a polaron at site $i$ with values either 0 or 1 and $V(i,j) = V_0 exp(-r_{ij}/r_s)/r_{ij}$ is the Yukawa potential that describes the screening. $V_0 = e^2/\epsilon_0 a$ in CGS units and $r_{ij} = |r_i - r_j|$, where $r_i$ is the dimensionless position of the $i$-th polaron and $r_s$ is the dimensionless screening radius. The value 1 for both $|r_i|$ and $r_s$ corresponds to one lattice constant $a$. $\epsilon_0$ is the static dielectric constant of the material. Polarons can only occupy the sites of the underlying triangular lattice with the lattice constant $a$ and the ratio of polarons in the system divided by the number of lattice sites is expressed as the filling $f$. The Monte-Carlo method used to simulate the model was described previously[24]. We studied the phase diagram of such a system at fixed values of $f$ and $r_s$. The CLG interpretation of polarons assumes a system of interacting phonons and repulsive electrons that is canonically transformed into a system of interacting small polarons in the strong electron-phonon coupling limit. We neglect spin effects, assume a screened Coulomb interaction, and assume that the hopping of polarons $\tilde{t} \ll V_0$, which is justified by the static nature of observed charges.

**Acknowledgements**
The work was supported by ERC ADG Trajectory (GA320602) and the Slovenian Research Agency (project P10040 and young researcher grants, P17589 and P08333). This project has received funding from the European Union's Horizon 2020 research and innovation program under the Marie Skłodowska-Curie grant agreement No 701647. We would like to thank Petra Šutar and Aleš Mrzel for sample growth.

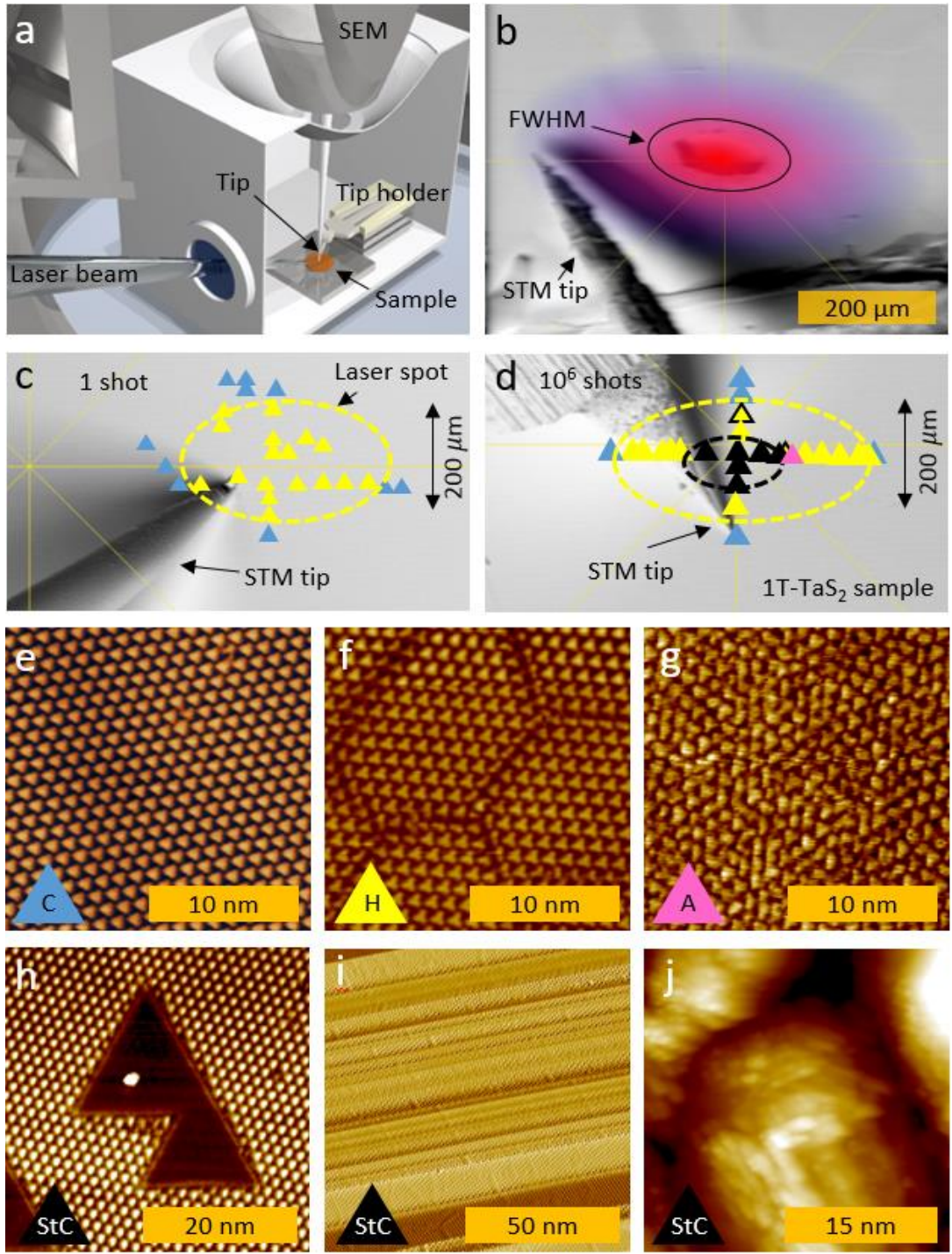

**Figure 1.** The STM experiment. a) A scheme of the optically excited STM setup. b) SEM image of an area exposed to 22 mJ/cm² peak fluence showing damage at the center of the laser spot, which was initially used to define the beam position. The local fluence decreases as we move away from the center of the beam (from red to blue as marked on the image). c) and d) SEM images of the sample with the marked state profile (color code from Figure 1 e)-j)) after excitation with a single shot and multiple shots respectively. e)-j) STM images of various states achieved by photoexcitation at 5K. e) ground C state, with a commensurate CDW and a perfect triangular lattice of polarons (bright spots), f) H state with CDW and domains of ordered polarons, g) A state with hyperuniform polaronic order, h) polytype transformation (at 77K), where dark regions represent the 1H polytype monolayer which has no charge modulation at this temperature[10] i) 1D surface ripples and j) sample melting.

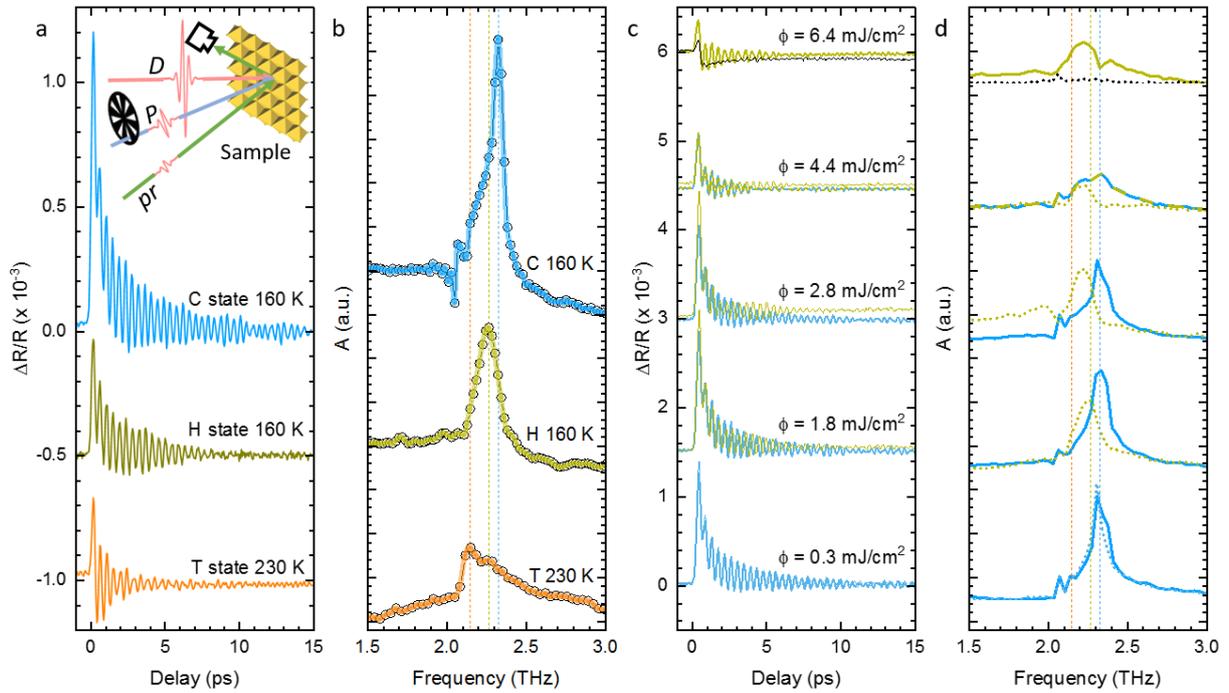

**Figure 2.** a) and b) The transient reflectivity oscillations and their Fourier spectra for C, H and T state, measured with three-pulse (C and H states) and two-pulse (T state) technique. The inset in a) shows the schematic of the setup. The vertical dashed lines in b) mark the peaks and are also shown in d) for comparison. c) Three-pulse D-P-Pr transient reflectivity measurements for different fluences at 30 ps (bold lines) and 1 ms (thin lines) showing different transition outcomes at 160 K , d) their respective Fourier transforms. The bold and dotted spectra correspond to the measurements at 30-ps and 1-ms, respectively.

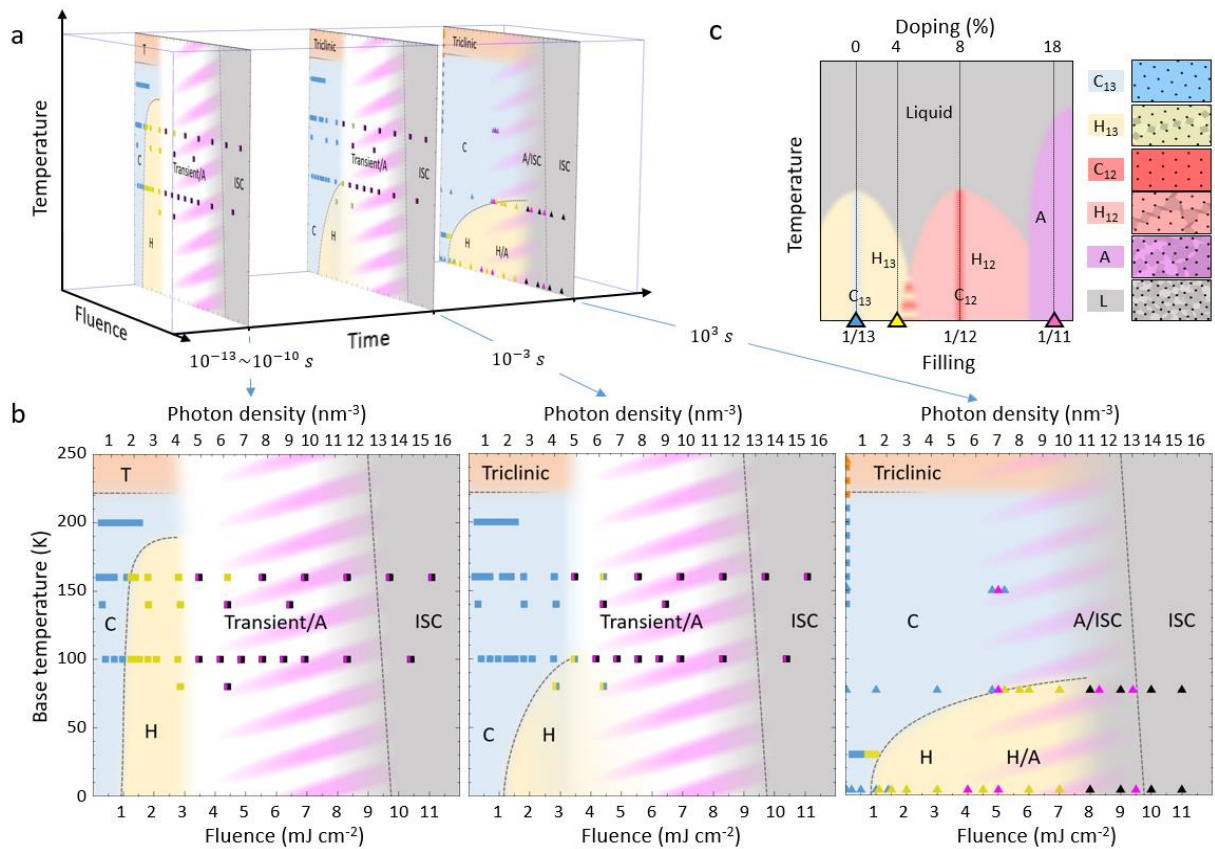

**Figure 3.** The time-domain phase diagram. a) Schematic timeline of the phase diagram evolution b) Detailed phase diagrams on the three timescales, as indicated. Triangles represent the STM data and squares represent the transient reflectivity data (color code from Figure 1 e)-j), and white represents the unresolved transient states). Note the significant shift of the long-time H state stability at high temperatures to higher fluences. c) The equilibrium phase diagram obtained from simulations[24]. The C phase (1/13 filling-blue), the H phase (around 1/13 filling-yellow) and the A phase (~1/11 filling-violet) are all observed experimentally. The predicted phases at and around 1/12 filling (red) are not observed in any of the experiments. The triangles and the dashed lines show at which filling values we observe the various states by STM.

# Supporting Information

**A time-domain phase diagram of metastable states in a charge ordered quantum material**

*Jan Ravnik, Michele Diego, Yaroslav Gerasimenko, Yevhenii Vaskivskyi, Igor Vaskivskyi, Tomaz Mertelj, Jaka Vodeb and Dragan Mihailovic*$^*$

**Details of STM experiments**

*Hidden state*

The threshold fluence for switching to the hidden state as seen with the STM appears to change with the temperature (Figure S1). At 4 K the average threshold fluence for observing the H state is 1.5 mJ/cm$^2$, while at 77 K we only observe the H state where the fluence is larger than 5.7 mJ/cm$^2$ (the observed area of the H state on the sample is smaller for the same laser beam). This is due to the faster relaxation of the H state at higher temperatures, which happens on the outskirts of the beam and prevents us from observing the whole switched area of H state with the relatively slow STM device. The observation could as well imply a higher switching threshold at higher temperatures, but the transient reflectivity measurements do not support this claim. Another important observation is that the observed switching fluence stays independent from the number of the incident pulses, both at 77 K and at 4 K. A very slight decrease in the threshold fluence (increase of the switched area) at 4 K with increasing the number of pulses is most likely the consequence of the spatial drift of the beam during long exposures. We conclude that the laser induced accumulated heating and nonequilibrium photoexcitations are well-separated effects.

*Amorphous state*

The amorphous state does not appear consistently across all experiments. In some cases, it is observed in patches within a larger area of the H state, while in other cases it is mixed with ISC areas. At 4 K the amorphous state was found both surrounded by H state and mixed with ISC, while when excited at 77 K it was only found mixed with ISC. On heating the A state from 4 K above 77 K, it was found surrounded by the C state when not mixed by ISC. In contrast with the electric switching[1], we have not found a way to controllably use optics to reproducibly switch the sample to the A state. However, we have found the boundary conditions at which the A state is likely to appear in optical experiments.

*Irreversible structural changes*

We observe no correlation between the sample temperature and the area of the induced ISC at the sample temperatures of 4 and 77 K (Figure S2). However, the area of ISC increases with the number of laser pulses, implying that at large fluences the accumulated heating plays a more critical role as opposed to low fluences. In most cases the threshold fluence to induce ISC is between 7 and 10 mJ/cm$^2$. This is particularly important for the polytype transformations[2], which were never observed in a single shot experiment, as they are a consequence of heating the sample, rather than an ultrafast electronic process. The same experimental conditions (power, exposure time and temperature) that lead to the polytype transformation in one case can create a large visible hole in the sample in another case. Single shot experiments can also have different ISC outcomes. When switching the sample with a single shot at two well-separated positions on the same crystal, we did not observe any ISC in one case, while in the other case we were able to see ISC in the form of melted sample in the middle. In both experiments the peak fluence exceeded 10 mJ/cm$^2$ and the sample was held at 4 K. This clearly suggests that the outcomes at high fluences are strongly dependent on local properties of the sample, which are responsible for uneven sample cooling. The heat transfer is determined by the quality of the thermal coupling between the top layer(s) and the bulk sample and/or the sample holder. This varies with the amount of glue, the thickness of the sample and the interlayer structure, which are not very well controlled experimental variables.

*Boundaries between the photoinduced states*

While the H state is usually found in large homogeneous patches (> 1 µm$^2$), the A state and ISC more likely appear in smaller patches intermixed with other states. The boundaries between the states are usually gradual. As an example, a transition between an H area and a completely melted sample area generally includes a few intermediate areas. First, transiting from the pure H state, we see the H state gets mixed with A state areas or covered with lots of small (< 5 nm) material deposits when moving towards the higher fluence region. Further, the ratio of the H area and the melted parts changes and larger melted parts start appearing more frequently (Figure S3b and c) until the H parts completely disappear in favor of the melted region. Similar kind of transitions are observed in all cases, many times including spots where more than two states are intermixed (Figure S3a).

As a contrast to Figure 1h in the main text, we show Figure S3d, where most of the area is transformed to 1H polytype with only small triangles of 1T.

**Transient reflectivity measurements of switching**

*Transient reflectivity at 100 and 160 K*

The first set of D-P-p measurements was done at 100 K, with the D-P delays of 400 ps, 30 ps, -3 ps and -30 ps and with the fluences ranging from 0.4 mJ cm$^{-2}$ to 10 mJ/cm$^2$ (Figure S4). In case where the D pulse arrives before the P-p sequence, we performed two separate sets of measurements, for the D-P delay of 30 ps and 400 ps. This way make sure that the observed state is not changing on the timescale of the measured P-p trace (30 ps). Since the excited states at $t_{D-P} = 30$ ps and $t_{D-P} = 400$ ps show the same characteristics, we conclude that they are stable on the 10-100 ps timescale and further consider only the measurements at $t_{D-P} = 30$ ps.

One can notice the generally decreasing amplitude of the oscillations when increasing the D pulse fluence. Comparing the measurements of the millisecond state to picosecond state, we can see that the oscillation amplitude is at all fluences at least partly restored, suggesting that some sort of relaxation takes place even at the highest fluences. At fluences higher than 10 mJ/cm$^2$ a visible damage (a hole) was created and we were unable to perform any further measurements on that part of the sample. From the measurements, where the D pulse hits the sample during the P-Pr sequence ($t_{D-P}$ = -3 ps) we see that above a certain threshold fluence the slow transient reflectivity component (neglecting the oscillatory part) changes to another value. This is associated with the switching to the H state and was described in details previously[3].

To more accurately determine the states, we did fast Fourier transforms (FFT) of the data for the millisecond and picosecond states (at $t_{D-P}$ = -30 ps (corresponding to the relaxed state after 1 ms) and 30 ps, respectively), which are shown in Figure S5. The maximum of the amplitude mode (AM) peak in the millisecond state stays approximately constant for the fluences of up to 3 mJ/cm$^2$, and corresponds to the C state. Above that fluence the peak becomes broader and weaker, suggesting a change which does not relax back to the initial state within 1 ms. The data for the picosecond state shows a similar change above 3 mJ cm$^{-2}$ as for the millisecond state, but the peaks are even more suppressed than for the millisecond state. The most important thing we see in the picosecond state is that at the fluences above 1 mJ cm$^{-2}$ the AM peak shifts to lower frequencies, indicating the switching to the H state.

An equivalent set of measurements was done also at 160 K. The data shows the same characteristics as at 100 K.

*Transient reflectivity at 200 K*

At 200 K, 1*T*-TaS$_2$ can be found in two different thermodynamically stable states, namely the C and NC state, depending on how the sample reached this temperature. Upon cooling it undergoes a phase transition from the NC to C state at 180 K, thus it is in the NC state at 200 K. Upon heating it stays in the C state until 220 K before reaching the triclinic state, thus it is in the C state at 200 K. By performing the experiment at 200 K we check if the switching to the H state can be distinguished from the effects of the accumulated heating of the sample to the triclinic or NC state by the train of D pulses. If the main mechanism is indeed heating, the sample should stay in the NC/triclinic state also after cooling and would not relax back to the C state at this temperature. In this case we would observe the characteristic oscillations of the triclinic or NC state when photoexciting the sample. Another thing that we could expect if heating was the transition mechanism, would be a decrease of the threshold fluence at higher temperatures. The highest measured fluences (up to 1.5 mJ/cm$^2$) are comparable or higher than the switching fluences observed at lower temperatures, but no change in the transient reflectivity is observed (Figure S6). We conclude that switching to the H state is a different process than supercooling a thermodynamically stable state.

*Transient reflectivity at 80 K*

We performed two-pulse and three-pulse measurements at 80 K, with two different D pulse fluences, in the range where we expected the switching to the H state to occur (2.86 mJ/cm$^2$) and at a bit higher value (4.41 mJ/cm$^2$), shown in Figure S7. At this temperature, the relaxation time of the H state should be comparable to the time between the laser pulses[4] and thus a two pulse measurement of the C state was also made for the comparison. By exciting the sample with a 2.86 mJ/cm$^2$ laser pulse, the AM peak moves to a lower frequency with respect to the AM peak in the C state, consistently with observations at other temperatures. When observing the millisecond state, the AM peak seems to be a combination of both C and H AM peaks, suggesting that the sample is only partially relaxed. This is consistent with the STM measurements at 77 K, where the sample partially relaxes. By increasing the fluence to 4.41 mJ/cm$^2$, we observe that the oscillations in the picosecond state have smaller amplitude consistently with the measurements at higher temperature.

**Theoretical phase diagram**

The parameter range of the filling $f$, relevant for comparison with the experiment is from 1/14 to 1/11, where all the experimentally observed states lie (Figure 3c). The model predicts a polaronic crystalline state (commensurate state) at $f = 1/12, 1/13$. Only 1/13 is observed experimentally and it corresponds to the thermodynamically stable C state. Upon increasing the value of $f$ from 1/13, a domain state (experimentally observed H state) emerges, where domains of the 1/13 lattice are separated by domain walls. Further increasing of $f$ introduces 1/12 lattice domains into the domain structure, which are not observed experimentally. The 1/13 domains disappear at some point and only 1/12 domains are left. When $f = 1/12$ is reached, polarons assume a perfect lattice pattern. Increasing $f$ from 1/12, a domain state with 1/12 lattice domains separated by domain walls reemerges, which eventually disappears and we are left with an amorphous state consisting of mixed striped states[5]. In the experiment, the only values of $f$ observed are 1/13 in the commensurate state, ~1/12.6 in the hidden state and ~1/11 in the amorphous state. To map the model to the experiment, we count the number of polarons per unit area in STM images. We assume the zero doping at the filling level of 1/13, which corresponds to the equilibrium C state. The deviations from the equilibrium state are linked to photodoping of the system which changes the value of $f$. The absence of observation of all the possible fillings between the values 1/12.6 and 1/11, which includes the 1/12 lattice as well as the mixture of 1/12 and 1/13 lattices is most likely due to various effects, which are not included in the model. The model neglects Fermi surface nesting, electron-phonon coupling, overlapping polaron deformations and itinerant carriers, which are all a part of the complex nature of 1$T$-TaS$_2$. As an example, Fermi surface nesting is the dominant factor in determining the high temperature incommensurate charge density wave state[6] and could as such significantly influence the outcome of the model. On the other hand, our correlated polaronic approach is valid to a certain degree due to its ability to predict the many different states, which could arise in other similar systems[5].

**Supplementary Figures**

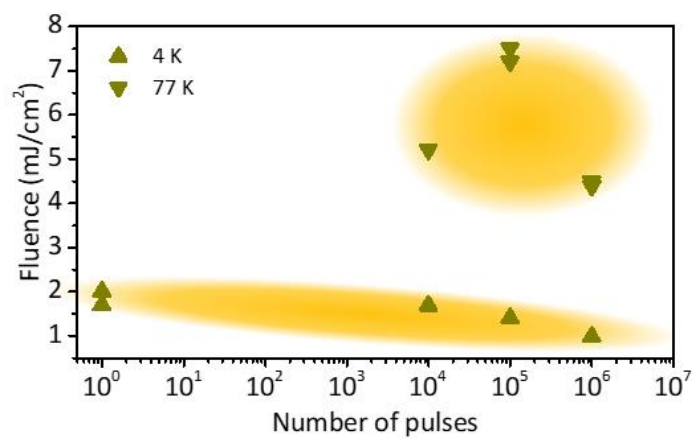

**Figure S1.** Threshold fluence for switching to the hidden state with respect to the number of shots. The upward and downward facing triangles show measurements at 4 K and 77 K respectively. The threshold fluence is lower at 4 K in all cases, which we attribute to the faster relaxation at 77 K. The yellow ellipses serve only as a guide to the eye.

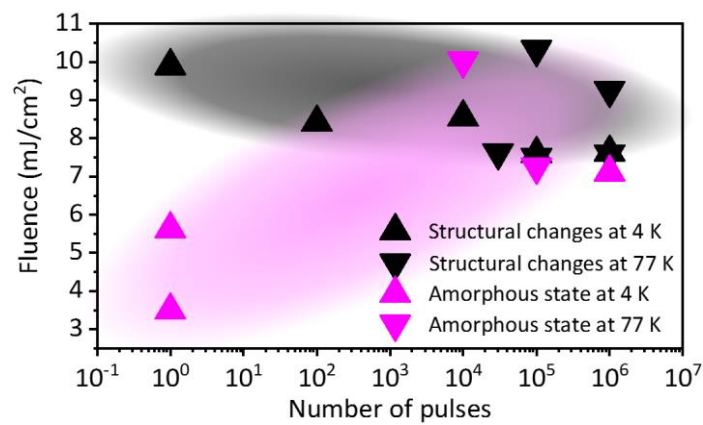

**Figure S2.** The observed threshold fluences for switching to the amorphous state (pink) and to induce structural changes (black) with respect to the number of shots. The upward and downward facing triangles show measurements at 4 K and 77 K respectively. The black and pink ellipses serve only as a guide to the eye.

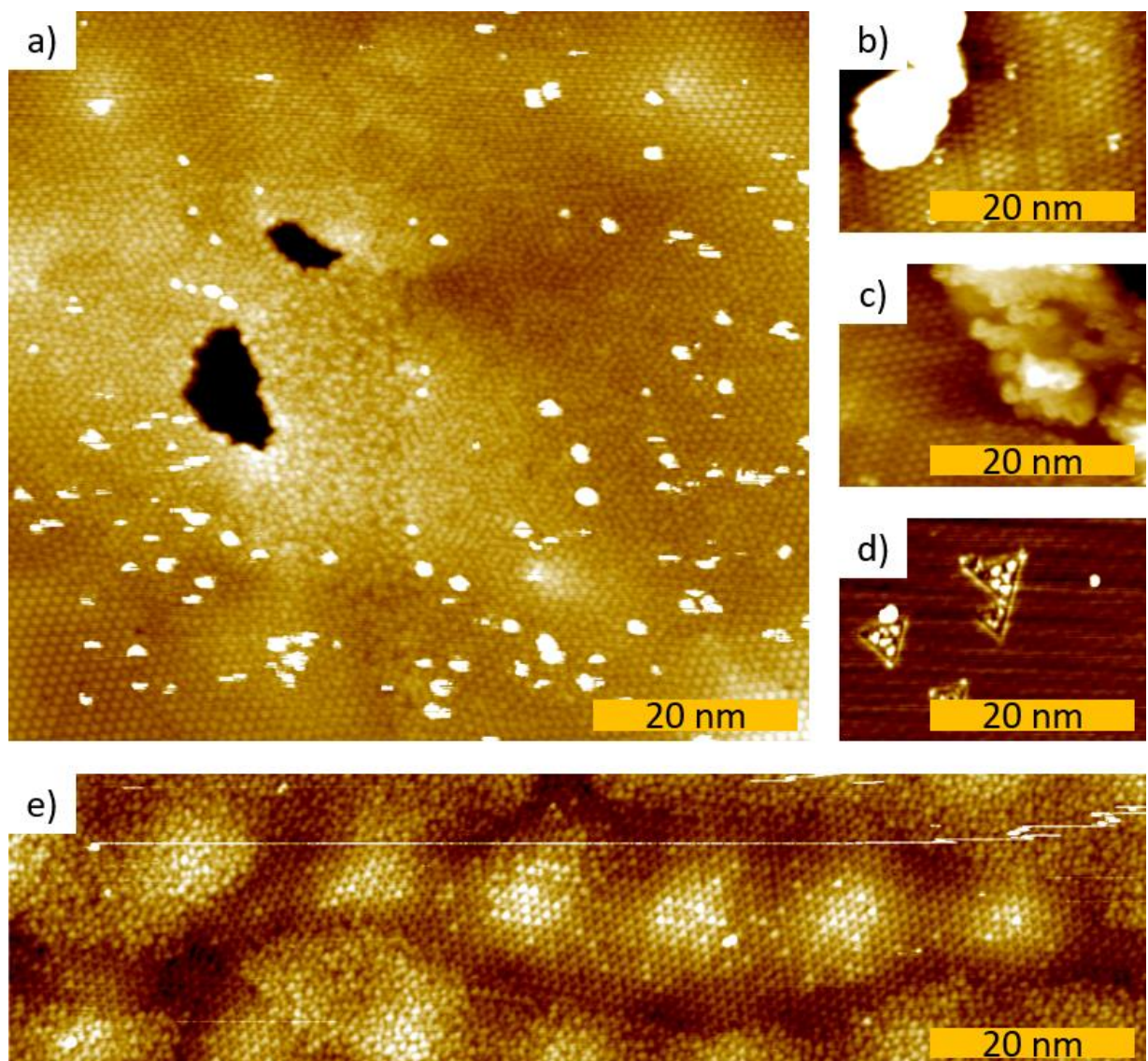

**Figure S3.** STM images of areas with more than one state present. a) Mixture of H and A state with ISC in form of material deposits and holes, b) and c) H state with large material deposits, d) 1H polytype area with small triangles of 1T polytype. e) a large A area encapsulating a small H area.

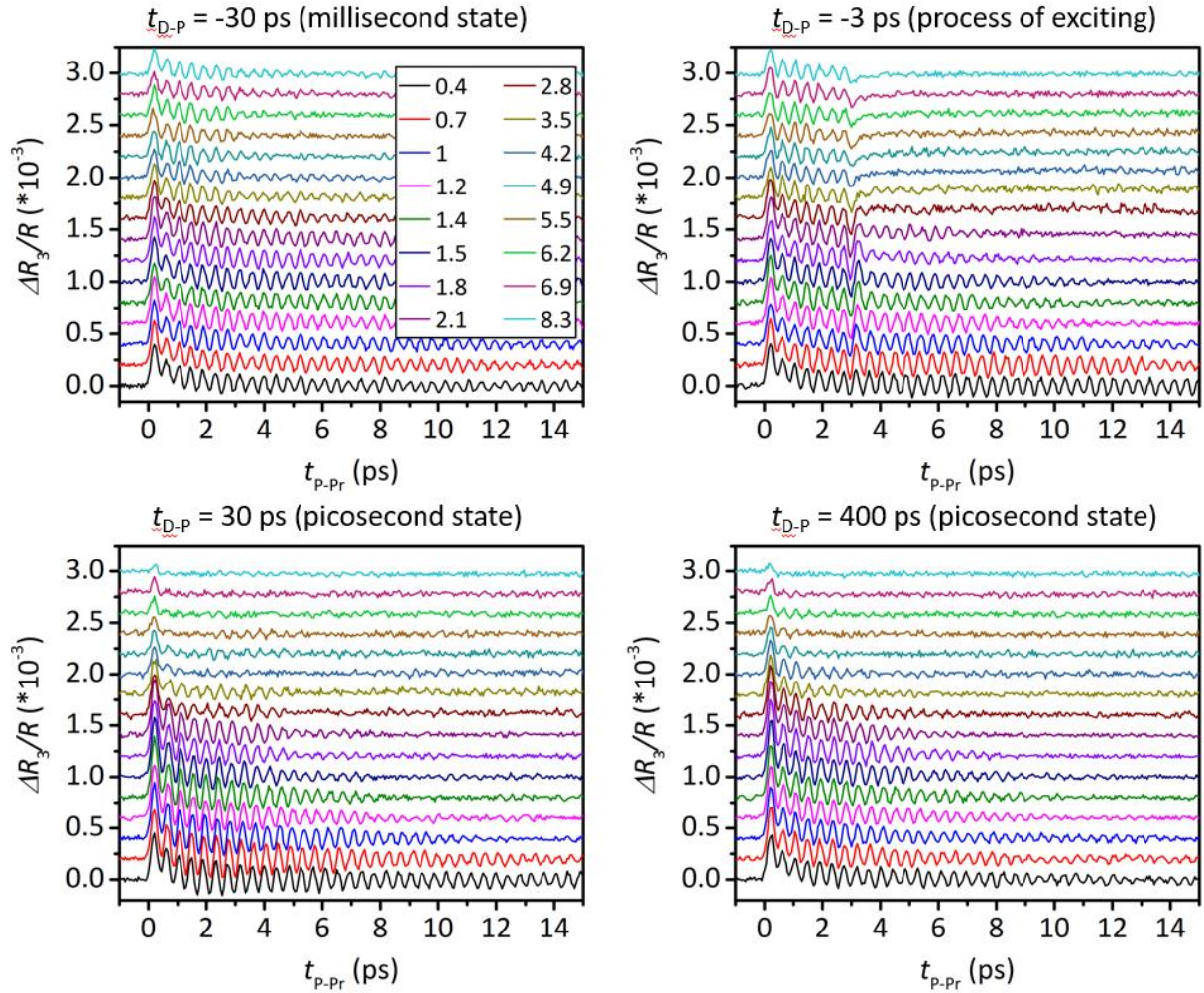

**Figure S4.** Three pulse transient reflectivity data at various D pulse fluences at T = 160 K. The measurements at $t_{D-P}$ = -30 ps show the sample in a 1 ms state, the measurements at $t_{D-P}$ = -3 ps show the arrival of D pulse during the P-Pr sequence and measurements at $t_{D-P}$ = 30 ps and $t_{D-P}$ = 400 ps show excited sample. We see that the latter two sets show identical behavior. The numbers in the legend represent fluences in mJ/cm$^2$.

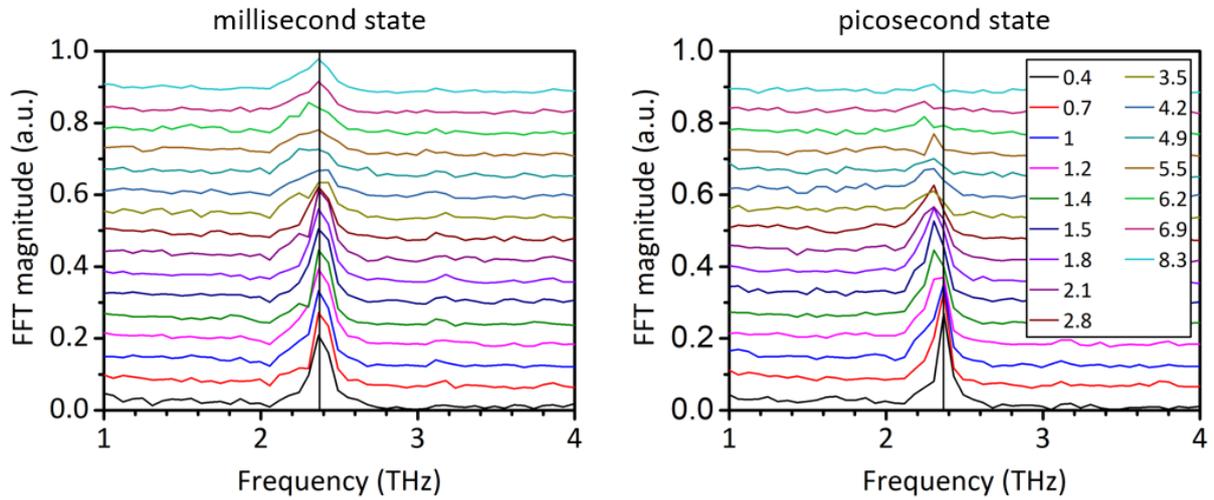

**Figure S5.** AM spectra of the data from Figure S4. The data of the millisecond state shows complete relaxation to C state up to the fluence of about 3 mJ/cm$^2$ and an obvious suppression of the AM peak at higher fluences. The data on the picosecond state shows switching to the H state at the fluence above 1 mJ/cm$^2$ and a lowering of the peak at fluences above 3 mJ/cm$^2$. The numbers in the legend represent fluences in mJ/cm$^2$.

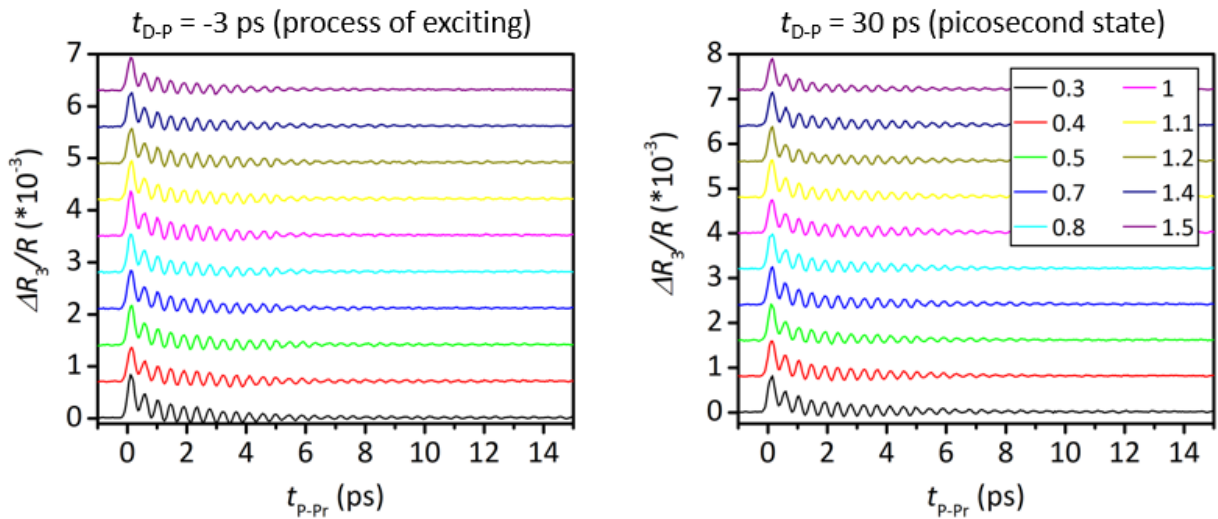

**Figure S6.** Three pulse transient reflectivity for various D pulse fluences at T = 200 K. The graphs show the measurements where the D pulse excites the sample during the P-p measurement ($t_{D-P}$ = -3 ps) and 400 ps before the measurement. The numbers in the legend represent fluences in mJ/cm$^2$.

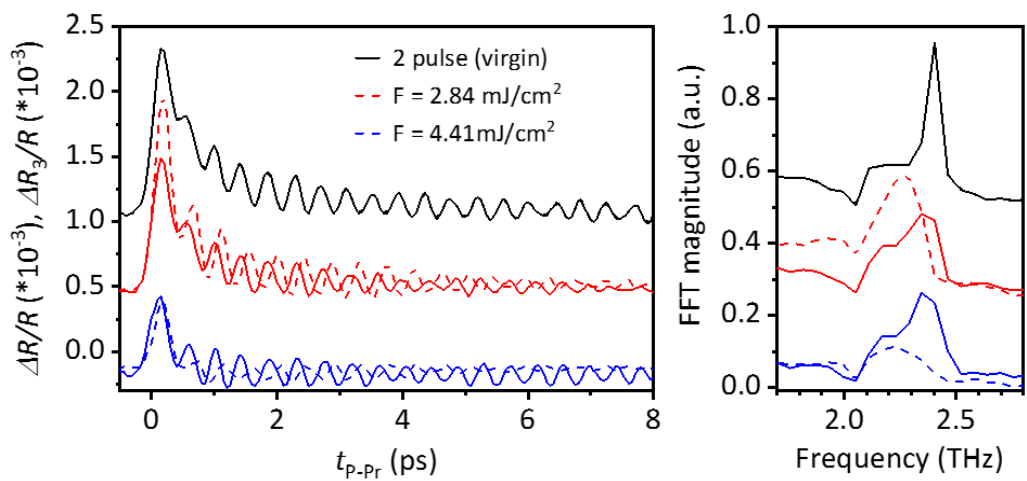

**Figure S7.** Two and three pulse pump probe measurements at 80 K (left) and their respective FFTs (right). The fluence of the D pulse was set to 0 mJ/cm$^2$ for the two pulse measurement (black) and 2.84 mJ/cm$^2$ (red) and 4.41 mJ/cm$^2$ (blue) in three pulse experiments. In the three pulse experiments the dashed and full lines represent the picosecond and millisecond states, respectively. We see that the sample does not relax from H state completely in 1 ms at 80 K. The amplitude of oscillations in the picosecond state significantly drops when the fluence is increased.